\documentclass[12pt]{article}
\usepackage{amssymb}
\usepackage{amsmath}
\usepackage{amstext}
\usepackage{graphicx,epsfig}
\usepackage{epsfig}
\usepackage{verbatim}
\usepackage{fancybox}
\usepackage{color}
\usepackage{ulem}
\usepackage{enumitem}
\usepackage{subfigure}
\usepackage{bbm}
\usepackage{parskip}
\usepackage{cite}
\usepackage{youngtab}

\newcommand{\Comment}[1]{{}}
\definecolor{MyDarkBlue}{rgb}{0.15,0.15,0.45}
\usepackage[linktocpage=true]{hyperref}
\hypersetup{
colorlinks=true,
citecolor=MyDarkBlue,
linkcolor=MyDarkBlue,
urlcolor=MyDarkBlue,
pdfauthor={Kurt Hinterbichler},
pdftitle={},
pdfsubject={hep-th}
}

\newcommand{\be}{\begin{equation}}
\newcommand{\ee}{\end{equation}}
\newcommand{\bal}{\begin{align}}
\newcommand{\eal}{\end{align}}
\newcommand{\bals}{\begin{align*}}
\newcommand{\eals}{\end{align*}}
\newcommand{\bea}{\begin{eqnarray}}
\newcommand{\eea}{\end{eqnarray}}
\newcommand{\beas}{\begin{eqnarray*}}
\newcommand{\eeas}{\end{eqnarray*}}
\newcommand{\nn}{\nonumber}

\def\({\left(}
\def\){\right)}

\numberwithin{equation}{section}

\begin{document}


\begin{center}
{\LARGE {Transverse diffeomorphism and Weyl invariant massive spin 2: \\ Linear theory  \\ \vspace{.2cm} }}
\end{center}
\vspace{2truecm}
\thispagestyle{empty} \centerline{
{\large {James Bonifacio,}}$^{a,}$\footnote{E-mail: \Comment{\href{mailto:james.bonifacio@physics.ox.ac.uk}}{\tt james.bonifacio@physics.ox.ac.uk}}
{\large {Pedro G. Ferreira,}}$^{b,}$\footnote{E-mail: \Comment{\href{mailto:p.ferreira1@physics.ox.ac.uk}}{\tt p.ferreira1@physics.ox.ac.uk}}
{\large {Kurt Hinterbichler}}$^{c,}$\footnote{E-mail: \Comment{\href{mailto:khinterbichler@perimeterinstitute.ca}}{\tt khinterbichler@perimeterinstitute.ca}}
}

\vspace{1cm}

\centerline{{\it
$^{a}$Theoretical Physics, University of Oxford,}}
\centerline{{\it DWB, Keble Road, Oxford, OX1 3NP, UK}}

\centerline{{\it
$^{b}$Astrophysics, University of Oxford,}}
\centerline{{\it DWB, Keble Road, Oxford, OX1 3RH, UK}}

\centerline{{\it
$^{c}$Perimeter Institute for Theoretical Physics,}}
\centerline{{\it 31 Caroline St. N, Waterloo, Ontario, Canada, N2L 2Y5 }}

\begin{abstract}

We use Kaluza-Klein reduction to construct flat-space massive spin-2 Lagrangians based on a kinetic term that has local Weyl and transverse-diffeomorphism gauge symmetries in the massless limit. This yields Lagrangians describing a free massive spin 2 which differ from the usual Fierz-Pauli theory, but are physically equivalent to it.   These Lagrangians require the use of auxiliary fields, which appear naturally from the higher-dimensional construction.  We discuss how these Lagrangians are related to each other and to the Fierz-Pauli theory through St\"uckelberg transformations, gauge fixings, and field replacements, and we use this to generalize the construction to spin-2 fields on Einstein manifolds. 

\end{abstract}


\newpage

\tableofcontents

\section{Introduction}
\parskip=5pt
\normalsize

A Lagrangian for a massive spin-2 field in four dimensions was first constructed by Fierz and Pauli \cite{Fierz:1939ix}. As pointed out in that paper, setting the mass to zero in their theory gives the Lagrangian for the 2 propagating degrees of freedom of a massless spin-2 field in four-dimensional Minkowski space. This massless theory is the linearization of the Einstein-Hilbert action about a flat background, and its Lagrangian possesses linearized diffeomorphisms as a gauge symmetry.

However, linearized general relativity (GR) is not the unique way to describe the helicity states of a free massless spin-2 field using an unconstrained symmetric rank-2 tensor. There is an alternative Lagrangian, sometimes called WTDiff,\footnote{The first mention of this Lagrangian that we know of is in \cite{Heiderich:1990bb}, whereas the name ``WTDiff," as well as the first extensive discussion, seem to appear in \cite{Alvarez:2006uu}; see also \cite{Blas:2007pp, Blas:2008uz, Alvarez:2010cg, Alvarez:2012uz}. For related earlier work, see \cite{Weyl:1944ve, VanNieuwenhuizen:1973fi, vanderBij:1981ym}.} that has as gauge symmetries linearized local Weyl transformations and linearized volume preserving diffeomorphisms.\footnote{Analogous statements also apply for all the higher spins \cite{Skvortsov:2007kz}.} However, unlike in the Fierz-Pauli theory, there is no naive mass term that can be added to the WTDiff Lagrangian to give a theory that propagates the 5 degrees of freedom of a massive spin-2 field in four dimensions---any such term gives a theory with 6 degrees of freedom, even the mass term consistent with the Weyl symmetry \cite{Alvarez:2006uu}.

In the many years following the Fierz-Pauli construction, attempts were made to generalize their theory by adding self-interactions to give a nonlinear theory of massive gravity. This was long believed to be impossible due to the generic presence of an extra degree of freedom (known as the Boulware-Deser ghost \cite{Boulware:1973my}). However, recently a nonlinear ghost-free generalization of the Fierz-Pauli action has been found \cite{deRham:2010ik,deRham:2010kj},  
thus initiating a recent surge of interest in massive gravity (see \cite{Hinterbichler:2011tt,deRham:2014zqa} for reviews). More specifically, there has been some interest in finding the possible spin-2 kinetic terms on which one can build such nonlinear theories \cite{Folkerts:2011ev,Hinterbichler:2013eza,Kimura:2013ika,deRham:2013tfa,Gao:2014jja,Noller:2014ioa}.

In this paper we revisit the linear WTDiff theory and attempt to find massive theories based on this alternative kinetic term.  Part of the motivation for this is the fact that the interaction terms which one may find for a degree of freedom can depend on which Lagrangian and gauge symmetries one uses in the linear theory as a starting point.  A simple illustration is the following: a massless spin-0 degree of freedom in four dimensions can be described by either a scalar field, or, through dualization, by a 2-form gauge potential.  The former allows zero-derivative self-interactions which preserve the number of degrees of freedom (simply write any potential for the scalar), the latter does not\footnote{As another example, in higher dimensions, a massless spin 2 can be realized using a symmetric tensor and the standard Fierz-Pauli Lagrangian with linearized diff symmetry, or it can be realized through a dual graviton which uses an exotic mixed symmetry field.  The former allows nonlinear interaction terms (the Einstein-Hilbert Lagrangian) but the latter does not \cite{Bekaert:2002uh}.}\cite{Henneaux:1997ha,Henneaux:1999ma,Knaepen:1999zr}.
If one were only aware of the latter formulation, many possible nonlinear theories would be missed.  Thus, it is advantageous to know all the possible ways of realizing a desired set of degrees of freedom at the linear level.
 
Another motivation for considering massive spin-2 theories based on the locally Weyl invariant kinetic term is the connection to unimodular gravity and the concomitant desire to shed light on the coupling of gravity to vacuum energy \cite{Weinberg:1988cp,Weinberg:2000yb}.  The linear WTDiff theory is closely related to unimodular gravity \cite{Einstein:1916vd,Henneaux:1984ji,Buchmuller:1988wx,Buchmuller:1988yn,Henneaux:1989zc,Ellis:2010uc, Alvarez:2005iy, Alvarez:2007nn,Smolin:2009ti, Barcelo:2014mua, Padilla:2014yea}, which is gauge fixed GR plus an extra overall spacetime-independent global degree of freedom, enforced through the constraint that the determinant of the metric be fixed.  This forbids a cosmological constant term at the level of the action, although because of the Bianchi identities this reappears in the equations of motion as an arbitrary integration constant.  
In the massive case, a small graviton mass can allow for self-accelerating solutions \cite{deRham:2010tw,Koyama:2011xz,Nieuwenhuizen:2011sq,Chamseddine:2011bu,D'Amico:2011jj,Gumrukcuoglu:2011ew,Berezhiani:2011mt}, giving the possibility of a technically natural solution to the cosmological constant problem \cite{ArkaniHamed:2002sp,deRham:2012ew,deRham:2013qqa} if some mechanism for eliminating a large bare cosmological constant can be made to work \cite{Dvali:2002pe,ArkaniHamed:2002fu,Dvali:2007kt,Patil:2010mq}. 
 
In this paper, we reexamine the question of whether an action for a massive spin 2 based on the WTDiff kinetic term can be constructed.\footnote{See \cite{Dalmazi:2014gaa} for a Weyl invariant massive spin-2 theory that is described using a nonsymmetric rank-2 tensor.} We use a Kaluza-Klein (KK) reduction \cite{Kaluza:1921tu,Klein:1926tv} (see \cite{Rindani:1985pi,Rindani:1988gb,Aragone:1988yx, Hinterbichler:2013kwa} for generalities on the KK reduction of linear theories with gauge symmetry) to find the massive version of the massless WTDiff theory in $D$ dimensions by dimensionally reducing the $(D+1)$-dimensional massless theory on a circle and truncating to a single massive mode.  This yields a Lagrangian describing the degrees of freedom of a massive spin-2 particle which is based off of a Weyl invariant kinetic term.  However, unlike the Fierz-Pauli case, the extra-dimensional components of the higher-dimensional metric cannot be completely gauged away without reverting the structure of the kinetic term to its Fierz-Pauli form.  As a result, the theory contains a nondynamical auxiliary field whose equations of motion combine with those of the tensor to enforce the constraints appropriate for a propagating massive spin 2. (In fact, such an auxiliary field was present already in the original construction of Fierz and Pauli, where the spin-2 degrees of freedom were embedded in a symmetric-traceless tensor and an auxiliary scalar helped to enforce a constraint.)

Both the Fierz-Pauli and WTDiff theories propagate the same degrees of freedom, although with different gauge groups in the massless case.  The two theories are not directly related by a field redefinition, but they can be realized as different gauge fixings of a parent theory that has a larger gauge group containing both smaller ones. In the massless case, this parent theory is the linearization of the theory of a conformally coupled scalar \cite{Deser:1970hs} and has both unconstrained diffeomorphisms and Weyl transformations as gauge symmetries.\footnote{As we discuss in Sec. \ref{section:cst} (at the linear level), the conformally coupled scalar theory can be obtained from the Einstein-Hilbert Lagrangian through a St\"uckelberg transformation. The resultant local Weyl symmetry leads to no nontrivial physical currents, as recently pointed out in \cite{Jackiw:2014koa}. Indeed, this is true of all symmetries that arise from St\"uckelberg transformations.  The utility of the St\"uckelberg trick is not to argue for physical currents, but as a tool to make manifest certain properties and equivalences of effective field theories (such as strong coupling scales and possible interactions).} From this theory, we can gauge fix to recover either the massless Fierz-Pauli theory or the massless WTDiff theory. Using KK reductions we generalize this picture to the massive theories, and in particular we construct the massive version of the (linearization of the) conformally coupled scalar theory.
This picture of gauge fixing a parent theory also gives a quick method for obtaining the WTDiff-invariant theories from their fully diff-invariant siblings through a noninvertible field replacement, which avoids the KK reduction. We use this to write down the curved-space versions of the above theories, in the case where the background is an Einstein space.  We pay particular attention to the formulation of the partially massless theories in this setup.

\textbf{Conventions:}  We use the mostly plus signature and freely integrate by parts in Lagrangians. $D$ is the number of spacetime dimensions, which we will keep arbitrary, but because both standard gravity and massive gravity have no local degrees of freedom in $D=2$, we consider only $D \geq 3$. 

\section{Flat-space massless theories} \label{sec:flatmassless}
\parskip=5pt
\normalsize

In this section we review the various Lagrangians which describe a massless spin 2 using a symmetric tensor, and discuss their relationship to each other through gauge fixings and field replacements, as summarized in Fig.~\ref{plot1}. 
The most general Lorentz invariant quadratic Lagrangian for a symmetric tensor field $h_{\mu\nu}$ on flat space, containing {exactly} two derivatives, is
\be {\cal L}_{{\rm FP},D}=a_1\,\partial_\lambda h_{\mu\nu}\partial^\lambda h^{\mu\nu}+a_2\,\partial_\lambda h_{\mu\nu}\partial^\mu h^{\lambda\nu}+a_3\,\partial_\mu h^{\mu\nu}\partial_\nu h+a_4\,\partial_\mu h \partial^\mu h, \label{laggeneral}
\ee
where the trace is denoted $h\equiv h^\mu_\mu$.
Among the four parameters $a_1,\ldots,a_4$ there are two redundancies: an overall scaling of $h_{\mu\nu}$ and the field redefinition $h_{\mu\nu}\rightarrow h_{\mu\nu}+c\, h \eta_{\mu\nu}$ for constant $c\not= {-1/ D}$ (ensuring invertibility).  This leaves a two-parameter family of possibilities. It turns out that there are precisely two choices of coefficients (up to the aforementioned redundancies) that lead to theories propagating precisely the degrees of freedom of a massless spin 2 and nothing more\footnote{{Adding a term $bh^2$, with arbitrary constant $b$, to these massless spin-2 theories does not change the number of local propagating degrees of freedom. However, this term reduces the size of the gauge group in both cases so that the kinetic terms are no longer completely fixed by the gauge symmetry, and introduces an arbitrary integration constant in the Fierz-Pauli theory.  We will restrict to the cases with maximal gauge symmetry, but it would be interesting to pursue the dimensional reduction of the more general possibilities.}} \cite{Alvarez:2006uu}.  As we now review, these theories called massless Fierz-Pauli and WTDiff are characterized by different gauge symmetries.  They can also be realized as different gauge fixings of another massless spin two theory, the linearized conformally coupled scalar-tensor theory. Finally, gauge fixing the trace to vanish in WTDiff gives the massless spin-2 theory in terms of a traceless tensor.

\subsection{Massless Fierz-Pauli}

The standard massless Fierz-Pauli Lagrangian in flat space is

\be {\cal L}_{{\rm FP},D}=-{1\over 2}\partial_\lambda h_{\mu\nu}\partial^\lambda h^{\mu\nu}+\partial_\lambda h_{\mu\nu}\partial^\mu h^{\lambda\nu}-\partial_\mu h^{\mu\nu}\partial_\nu h+{1\over 2}\partial_\mu h \partial^\mu h. \label{FPmassless}
\ee
This is the linearization of the Einstein-Hilbert Lagrangian, $\label{EH}
\mathcal{L} = \frac{1}{2 \kappa^2} \sqrt{-g} R,$ 
around a Minkowski background, $g_{\mu \nu} = \eta_{\mu \nu} +2 \kappa h_{\mu \nu}$.  The Lagrangian \eqref{FPmassless} is invariant under linearized diffeomorphisms 
\be \delta h_{\mu\nu}=\partial_\mu\xi_\nu+\partial_\nu\xi_\mu,\label{lindifff}\ee
for a vector gauge parameter $\xi_\mu$. The coefficients of \eqref{FPmassless}, up to an overall scaling of $h_{\mu\nu}$, are completely fixed by the gauge symmetry \eqref{lindifff}.

\subsection{Massless WTDiff}
\label{MasslessWTDiff}

The flat-space WTDiff theory is
\be {\cal L}_{{\rm WTDiff},D}=-{1\over 2}\partial_\lambda h_{\mu\nu}\partial^\lambda h^{\mu\nu}+\partial_\lambda h_{\mu\nu}\partial^\mu h^{\lambda\nu}-{2\over D}\partial_\mu h^{\mu\nu}\partial_\nu h+{D+2\over 2D^2}\partial_\mu h \partial^\mu h. \label{WTDiffmassless}
\ee
It is invariant under the gauge symmetries
\be \delta h_{\mu\nu}=\partial_\mu\xi_\nu^T+\partial_\nu\xi_\mu^T+\chi\eta_{\mu\nu},\ \ \ \ \  \partial^{\mu} \xi_\mu^T =0.  \label{gaugesymm1}\ee
Here $\chi$ is a scalar gauge parameter and $\xi_\mu^T$ is a vector gauge parameter which is required to be transverse: $\partial^{\mu} \xi_\mu^T =0$. 
We can think of $\chi$ as a linearized Weyl transformation, and $\xi_\mu^T$ as a linearized volume preserving diffeomorphism.  The coefficients in \eqref{WTDiffmassless} are completely fixed, up to overall scaling of $h_{\mu\nu}$, by demanding the gauge symmetry \eqref{gaugesymm1}.

While this Lagrangian describes $\frac{D(D-3)}{2}$ propagating local degrees of freedom, there is a slight subtlety in interpreting these as flat-space spin-2 helicity modes. This is because \eqref{WTDiffmassless} describes, in addition to the $\frac{D(D-3)}{2}$ helicity-2 degrees of freedom at each point, an extra global degree of freedom, in the sense that there is one extra global initial condition that needs to be specified to uniquely determine the dynamics. This shows up as an integration constant in the equations of motion and behaves as a tadpole term (a nonzero vacuum expectation value for $h_{\mu \nu}$).  This is the linear analogue of the cosmological constant appearing as an integration constant in unimodular gravity.

To see this, first use the $\chi$ gauge symmetry to set $h=0$; this leaves a residual transverse-diffeomorphism (TDiff) symmetry.  Then taking a divergence of the equations of motion we find $\partial_\rho(\partial^{\mu} \partial^{\nu} h_{\mu \nu})=0$, which can be integrated once to give
\be \label{eq:ree}
\partial^{\mu} \partial^{\nu} h_{\mu \nu} = c,
\ee
for some (gauge-invariant) constant $c$. 
{Integrating this again gives 
\be 
\partial^{\mu} h_{\mu \nu} = \frac{c}{D} x_{\nu} + a^T_{\nu}(x),
\ee
where $x^\mu$ are the spacetime coordinates and $a^T_{\mu}(x)$ is an arbitrary transverse vector field, $\partial^{\mu} a^T_{\mu}(x) =0$. This arbitrary transverse vector field can be gauged to zero using the residual TDiff symmetry.}
The equations of motion then give the unsourced equation $\Box h_{\mu \nu} =0$, but the tensor is not transverse. We can get a transverse tensor at the expense of a source by renaming $\bar{h}_{\mu \nu} = h_{\mu \nu} -\frac{c }{2D}x^2 \eta_{\mu \nu}$. This gives
\be \label{spin2sourced}
\Box \bar{h}_{\mu \nu} =-c \eta_{\mu \nu} \quad \text{and} \quad \partial^{\mu}  \bar{h}_{\mu \nu} = 0,
\ee
which shows that the integration constant $c$ plays the role of a tadpole. 
 
\subsection{Conformal scalar tensor} \label{section:cst}
${\cal L}_{{\rm FP},D}$, and formally ${\cal L}_{{\rm WTDiff},D}$, can be obtained by gauge fixing a parent scalar-tensor theory that has full diffeomorphism and Weyl gauge invariance, which is obtained by making a St\"uckelberg replacement (or ``St\"uckelberging")\footnote{The St\"uckelberg procedure is a method of introducing gauge invariance by including extra (St\"uckelberg) fields \cite{stuckelberg}. See, e.g., Sec. 4 of \cite{Hinterbichler:2011tt} for a review of this in the case of linear spin 1 and 2. } $h_{\mu\nu}\rightarrow h_{\mu\nu}+\varphi\,\eta_{\mu\nu}$ in the Fierz-Pauli theory \eqref{FPmassless}:
\begin{align} \label{conformal} {\cal L}_{{\rm Conformal},D}=& -{1\over 2}\partial_\lambda h_{\mu\nu}\partial^\lambda h^{\mu\nu}+\partial_\lambda h_{\mu\nu}\partial^\mu h^{\lambda\nu}-\partial_\mu h^{\mu\nu}\partial_\nu h+{1\over 2}\partial_\mu h \partial^\mu h \nn \\
&+(D-2)h^{\mu\nu}\left(\partial_\mu\partial_\nu\varphi-\eta_{\mu\nu}\square\varphi\right)+{1\over 2}(D-1)(D-2)(\partial\varphi)^2.
\end{align}
We now have unrestricted diffeomorphism symmetry and Weyl symmetry,
\be
\delta h_{\mu\nu}=\partial_\mu\xi_\nu+\partial_\nu\xi_\mu+\chi\,\eta_{\mu\nu},\ \ \ \ \delta\varphi=-\chi.
\ee
The theory \eqref{conformal} can also be obtained by linearizing the conformally coupled scalar-tensor theory \cite{Deser:1970hs}, $\label{conformalnonlinear}
\mathcal{L} =  \frac{1}{2} \sqrt{-g} \left( g^{\mu \nu} \partial_{\mu} \varphi \partial_{\nu} \varphi +  \frac{D-2}{4(D-1)} \varphi^2 R\right)$, around a Minkowski background with a nonzero constant scalar profile.

From the conformal scalar-tensor theory \eqref{conformal}, we get Fierz-Pauli \eqref{FPmassless} by fixing the gauge $\varphi=0$.  The residual gauge symmetry which preserves this gauge choice is precisely \eqref{lindifff}.  We would get WTDiff \eqref{WTDiffmassless} by fixing the gauge $\varphi=-{1\over D}h$, but this is not in fact a legal gauge fixing because the gauge cannot be reached.  The residual gauge symmetry which preserves this gauge choice is precisely \eqref{gaugesymm1}. 
Formally, WTDiff can be obtained from Fierz-Pauli by making the replacement $h_{\mu\nu}\rightarrow h_{\mu\nu}-{1\over D}h\eta_{\mu\nu}$, but this is not an invertible transformation.

\subsection{Traceless theory}

Finally, a massless graviton can be described by a Lagrangian for a traceless tensor $\tilde h_{\mu\nu}$,
\be {\cal L}_{{\rm traceless},D}=-{1\over 2}\partial_\lambda \tilde h_{\mu\nu} \partial^\lambda \tilde h^{\mu\nu \, }+\partial_\lambda \tilde h_{\mu\nu}\partial^\mu \tilde h^{\lambda\nu \, }, \label{tracelessm}
\ee
which has only TDiff gauge symmetry 
\be \delta \tilde h_{\mu\nu}=\partial_\mu\xi_\nu^T+\partial_\nu\xi_\mu^T,\ \ \ \ \  \partial^{\mu} \xi_\mu^T =0.  \label{gaugesymm1b}\ee
We obtain this theory from the WTDiff theory by using the Weyl symmetry to eliminate the trace of $h_{\mu\nu}$. 
Going the other way, we can obtain the WTDiff theory from the traceless theory by St\"uckelberging the trace:  $\tilde {h}_{\mu \nu} \rightarrow h_{\mu \nu} - \frac{1}{D} h \eta_{\mu \nu}.$ 
The traceless theory \eqref{tracelessm} can be formally obtained from Fierz-Pauli by setting $h=0$, but this is {\it not} an algebraic gauge fixing (such a fixing is not in general ``legal" to perform in the action) so we should think of it only as a shorthand way of transforming between different theories.

\begin{figure}[h!]
\begin{center}
\epsfig{file=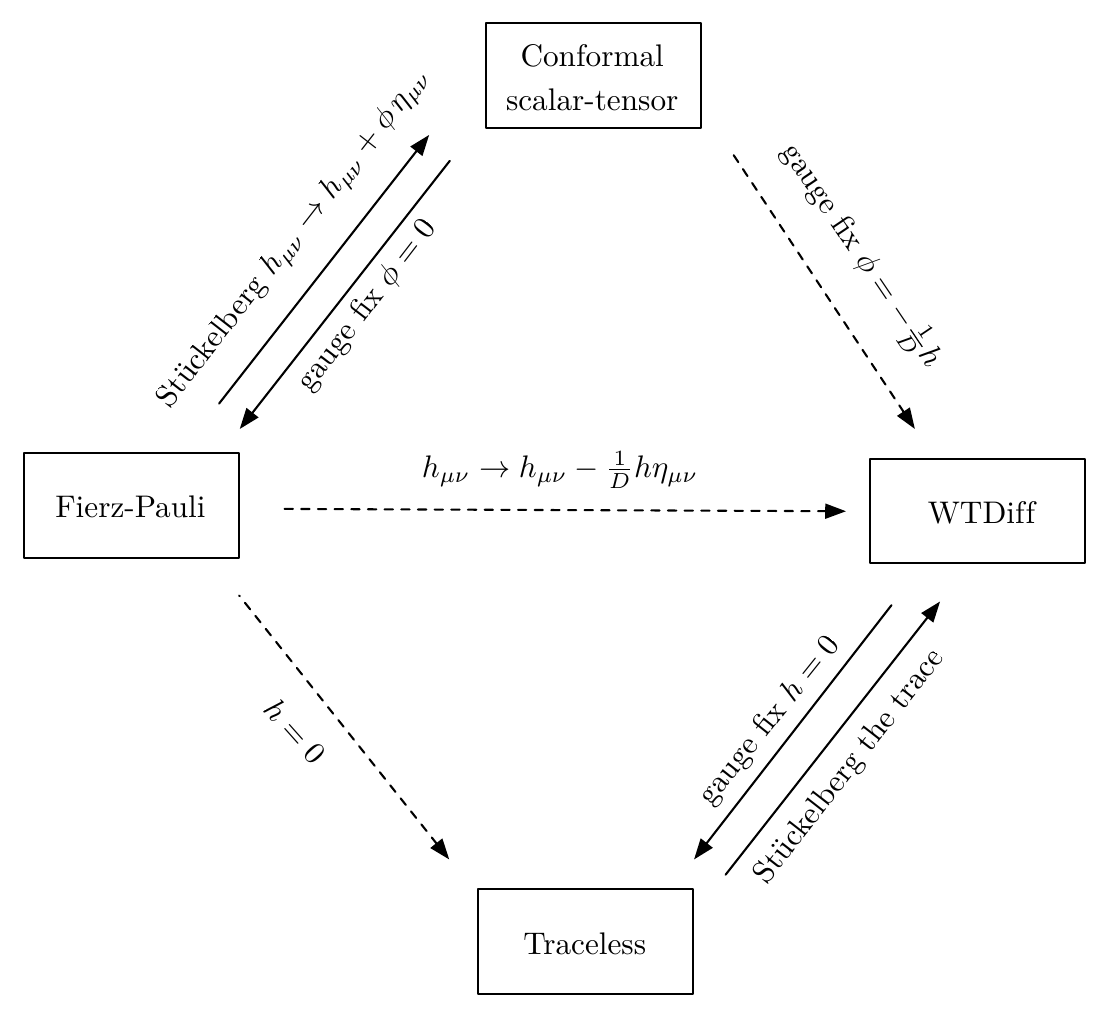,height=4.5in,width=5.0in}
\caption{Relations between massless actions. Solid lines are legal transformations, dotted lines are cheats.}
\label{plot1}
\end{center}
\end{figure}

The various massless theories and their relationships are illustrated in Fig. \ref{plot1}, where we show the transformations (both ``legal" and ``cheats") that can map between them.  We see that the WTDiff theory and the Fierz-Pauli theory cannot be reached from each other by legal transformations.  This is because they are not strictly equivalent; they are locally equivalent but they differ by a global degree of freedom.

\section{Flat-space massive theories} \label{sec:flatmassive}

The relativistic equations that describe the dynamics of the $\frac{(D+1)(D-2)}{2}$ polarizations of a massive spin-2 particle of mass $m$ are
\be \label{KG}
(\Box -m^2) {h}_{\mu \nu} =0, \ \ \ \ \partial^{\mu} {h}_{\mu \nu} =0,\ \ \ \ {h}_{} =0.
 \ee 
We look for modifications of the above Lagrangians involving mass terms for the tensor field that yield these equations and hence propagate exactly the $\frac{(D+1)(D-2)}{2}$ degrees of freedom of a massive spin 2. For the case of linearized GR, the answer---which has been known since the work of Fierz and Pauli---is to add the term $-{1\over 2}m^2(h_{\mu \nu}h^{\mu \nu} - h^2)$ to the massless action.\footnote{Fierz and Pauli first found the massive theory and then noted that this gave linearized GR in the massless limit.} The relative tuning between $h_{\mu \nu}^2$ and $h^2$ is needed to ensure that only 5 degrees of freedom propagate in four dimensions---modifying this tuning introduces a ghostlike extra degree of freedom. Now we can also try adding a generic quadratic mass term to the massless WTDiff Lagrangian, but in this case all such terms\footnote{{Except the term proportional to $h^2$, which gives a Lagrangian that propagates the correct number of massless degrees of freedom, as footnoted in Sec. \ref{sec:flatmassless}.}} give a Lagrangian that propagates 6 degrees of freedom, one of which is a ghost \cite{Alvarez:2006uu}. Thus, as noted by \cite{Alvarez:2006uu}, the WTDiff theory as compared to linearized GR appears more rigid against massive deformations.

\begin{figure}[h!]
\begin{center}
\epsfig{file=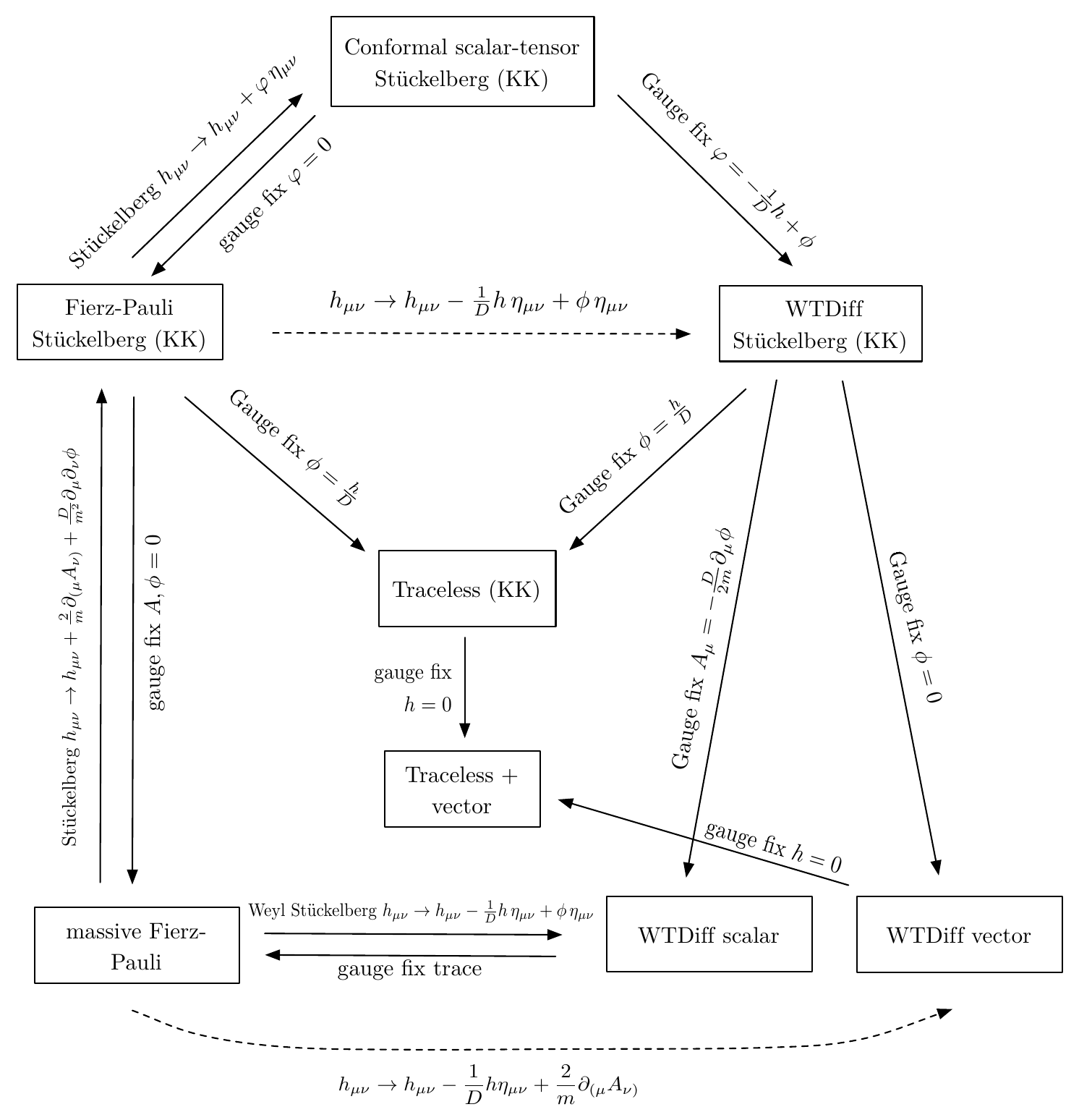,height=6.0in,width=5.5in}
\caption{The massive Lagrangians and some of their relationships. Solid lines are legal transformations, dotted lines are cheats.  The theories labeled with (KK) are the Kaluza-Klein reductions of the theories in Fig. \ref{plot1}, up to simple field redefinitions.}\label{massiverelationships}
\label{plot2}
\end{center}
\end{figure}

 As we show below, there is a massive spin-2 theory with a Weyl invariant kinetic term but this requires an auxiliary field, which is similar to the massive spin-2 theory as Fierz and Pauli originally wrote it. This theory can be obtained by performing a truncated KK reduction on the massless WTDiff theory, and is related to massive Fierz-Pauli by field redefinitions and gauge fixings. In what follows, we review how to find the massive Fierz-Pauli theory using KK reduction and then apply this technique to find the massive WTDiff theory and the massive conformally coupled scalar-tensor theory. {The Lagrangians we find and some relationships between them are summarized in Fig. \ref{plot2}, generalizing Fig. \ref{plot1} to nonzero mass.} 
 
\subsection{Massive Fierz-Pauli}
The Fierz-Pauli massive spin-2 theory, as usually written, is
\be
{\cal L}_{{\rm mFP},D}=-{1\over 2}\partial_\lambda h_{\mu\nu}\partial^\lambda h^{\mu\nu}+\partial_\lambda h_{\mu\nu}\partial^\mu h^{\lambda\nu}-\partial_\mu h^{\mu\nu}\partial_\nu h+{1\over 2}\partial_\mu h \partial^\mu h -\frac{1}{2}m^2(h_{\mu \nu}h^{\mu \nu} - h^2). \label{FPmassive}
\ee 
Note that putting $m=0$ gives back \eqref{FPmassless}, although there is no smooth $m \rightarrow 0$ limit since the number of degrees of freedom is different between the massless and massive theories\cite{vanDam:1970vg, Zakharov:1970cc}.

The massive Fierz-Pauli theory can be obtained by performing a KK reduction of the higher-dimensional massless Fierz-Pauli theory and then truncating to a single massive sector of the resultant infinite tower of fields. This is a consistent truncation at the linear level that we can derive automatically by replacing $H_{AB}$ in the higher-dimensional action with
\begin{align}  H_{\mu\nu}(x,y)& =\sqrt{2m\over \pi}\cos\left(my\right)h_{\mu\nu}(x) ,\\
H_{\mu y}(x,y)&=\sqrt{2m\over \pi}\sin\left(my\right)A_{\mu}(x) ,\\
H_{yy}(x,y)&=\sqrt{2m\over \pi}\cos\left(my\right)\phi_{}(x),
\end{align}
where capital latin indices are $(D+1)$-dimensional, greek indices are $D$-dimensional, and $y$ labels the extra compact dimension that we integrate over using the orthogonality relations 
\be
\int_0^{\pi\over m} \cos^2(my) \, dy =\int_0^{\pi\over m}\sin^2(my) \, dy = {\pi \over 2m}, \quad \int_0^{\pi\over m}\cos(my)\sin(my) \, dy=0.
\ee
Since the higher-dimensional graviton satisfies the higher-dimensional gauge transformation,
\be
\delta H_{AB} = \partial_{A} \Xi_{B} + \partial_{B} \Xi_{A},
\ee
we can get the lower-dimensional gauge transformations by also expressing $\Xi_A$ in terms of lower-dimensional Fourier modes,
\begin{align}
\Xi_{\mu}(x,y)&=\sqrt{2m\over \pi}\cos\left(my\right)\xi_{\mu}(x) ,\\
\Xi_{ y}(x,y)&=\sqrt{2m\over \pi}\sin\left(my\right)\xi_{}(x),
\end{align}
and equating components.

Reducing the $(D+1)$-dimensional massless Fierz-Pauli action in this way leads to
\be \label{mFPStueckelberg1}
{\cal L}_{{\rm FP},D+1}\rightarrow {\cal L}_{{\rm mFP},D} -\frac{1}{2} F_{\mu \nu}^2 -2m h^{\mu \nu}(\partial_{\mu} A_{\nu} - \eta_{\mu \nu} \partial^{\alpha} A_{\alpha})+ h^{\mu \nu}( \partial_{\mu} \partial_{\nu} \phi - \eta_{\mu \nu} \Box \phi).
\ee
\textcolor{black}{Redefining $\phi \rightarrow - D \phi$, to simplify later expressions, we end up with
\be \label{mFPStueckelberg}
 {\cal L}_{{\rm mFP},D} -\frac{1}{2} F_{\mu \nu}^2 -2m h^{\mu \nu}(\partial_{\mu} A_{\nu} - \eta_{\mu \nu} \partial^{\alpha} A_{\alpha})-D h^{\mu \nu}( \partial_{\mu} \partial_{\nu} \phi - \eta_{\mu \nu} \Box \phi),
\ee}
which has the gauge symmetries
\begin{align}\delta h_{\mu\nu}&=\partial_\mu\xi_\nu+\partial_\nu\xi_\mu \nn\\
 \delta A_\mu&=\partial_\mu\xi-m\xi_\mu,\nn\\
\delta\phi&=-\frac{2}{D}m\xi. 
\end{align}
This Lagrangian \textcolor{black}{labeled ``Fierz-Pauli St\"uckelberg" in Fig. \ref{plot2}} is just \eqref{FPmassive} after the St\"uckelberg procedure
\be
h_{\mu \nu} \rightarrow h_{\mu \nu} + \frac{1}{m}\partial_{\mu} A_{\nu} + \frac{1}{m}\partial_{\mu} A_{\nu} + \frac{D}{m^2}\partial_{\mu} \partial_{\nu} \phi.
\ee
In this case, $A_{\mu}$ and $\phi$ are both St\"uckelberg fields that can be gauge fixed to zero (unitary gauge), giving back \eqref{FPmassive}.

\subsection{Massive WTDiff}

Performing the KK reduction on the higher-dimensional WTDiff theory \eqref{WTDiffmassless} gives the massive WTDiff theory in the St\"uckelberg-ed form,
\begin{align} \label{fullmWTDiff} {\cal L}_{{\rm WTDiff},D+1}\rightarrow& {\cal L}_{{\rm WTDiff},D}+{2\over D(D+1)}\partial_\mu h^{\mu\nu}\partial_\nu h-{D^2+5D+2\over 2D^2(D+1)^2}\partial_\mu h \partial^\mu h-{1\over 2}m^2\left(h_{\mu\nu}^2-{D+3\over (D+1)^2}h^2\right) \nn\\
& -{1\over 2}F_{\mu\nu}^2-2mh^{\mu\nu}\left(\partial_\mu A_\nu-{2\over D+1}\eta_{\mu\nu}\partial\cdot A\right)+{2(D-1)\over D+1}m A^\mu\partial_\mu \phi \nn\\
& +{2\over D+1}h^{\mu\nu}\left(\partial_\mu\partial_\nu\phi-{D+3\over 2(D+1)}\eta_{\mu\nu}\square\phi\right)-{D-1\over (D+1)^2}m^2h\phi \nn \\
&-{(D-1)\over 2(D+1)^2}\left[(D+2)(\partial\phi)^2-{Dm^2}\phi^2\right]. \nn\\  
\end{align}
This action has the following gauge symmetries:
\begin{align}  \delta h_{\mu\nu}&=\partial_\mu\xi_\nu+\partial_\nu\xi_\mu+\chi\eta_{\mu\nu}, \nn\\
 \delta A_\mu&=\partial_\mu\xi-m\xi_\mu,\nn\\
\delta\phi&=\chi+2m\xi.
\end{align}
Here $\chi$ descends from the higher-dimensional Weyl gauge parameter and $\xi_\mu,\xi$ descend from the higher-dimensional diff parameter $\Xi^A$.  The transverse condition on the $(D+1)$-dimensional diff parameter, $\partial^A \Xi_A =0$, constrains the lower-dimensional gauge parameters to satisfy 
\be \label{Weylconstraint} \partial_\mu\xi^\mu+m\xi=0.\ee

\textcolor{black}{The Lagrangian \eqref{fullmWTDiff} can be simplified with the field redefinition
\begin{align} \phi& \rightarrow {1\over D}h-(D+1)\phi, \label{redefine1}
\end{align}
after which we end up with the St\"uckelberg-ed massive WTDiff theory, as represented in Fig. \ref{plot2},
\begin{align}
{\cal L}_{{\rm mWTDiff},D}(h, A, \phi) =& {\cal L}_{{\rm WTDiff},D}-{1\over 2}m^2\left(h_{\mu\nu}^2-{1\over D}h^2\right) \nn\\
& -{1\over 2}F_{\mu\nu}^2-2mh^{\mu\nu}\left(\partial_\mu A_\nu-{1\over D}\eta_{\mu\nu}\partial\cdot A\right)-2(D-1)m A^\mu\partial_\mu \phi \nn\\
& -2h^{\mu\nu}\left(\partial_\mu\partial_\nu\phi-{1\over D}\eta_{\mu\nu}\square\phi\right)-{(D-1)\over 2 }\left[(D+2)(\partial\phi)^2-{Dm^2}\phi^2\right], \nn\\ \label{mWTDiffminvector1}
\end{align}
which has the gauge symmetries
\begin{align} \delta h_{\mu\nu}&=\partial_\mu\xi_\nu+\partial_\nu\xi_\mu+\chi\eta_{\mu\nu}, \nn\\
 \delta A_\mu&=\partial_\mu\xi-m\xi_\mu,\nn\\
 \delta\phi&=-\frac{2}{D}m\xi,
\end{align}
and the constraint 
\be \label{Weylconstraint2} \partial_\mu\xi^\mu+m\xi=0.\ee}

Both $A_{\mu}$ and $\phi$ are St\"uckelberg fields, but---unlike in the massive Fierz-Pauli case---we cannot eliminate both $\phi$ and $A_{\mu}$ simultaneously, because of the constraint \eqref{Weylconstraint2}.  Thus there is no analogue of the pure unitary gauge found for the Fierz-Pauli theory.  The best we can do is to eliminate either $\phi$ or $A_{\mu}$.

\subsubsection{Gauge $\phi=0$}
Eliminating $\phi$ from \eqref{mWTDiffminvector1}, we end up with a tensor-vector description of the massive spin-2 field, whose massless part is the WTDiff Lagrangian and a free Maxwell Lagrangian. To do this, we use $\xi$ to set $\phi=0$. This gives
\be \label{mWTDiffvector}
{\cal L}_{{\rm mWTDiff},D}(h, A)= {\cal L}_{{\rm WTDiff},D}-{1\over 2}m^2\left(h_{\mu\nu}^2-{1\over D}h^2\right) -{1\over 2}F_{\mu\nu}^2-2mh^{\mu\nu}\left(\partial_\mu A_\nu-{1\over D}\eta_{\mu\nu}\partial\cdot A\right),
\ee
which has the residual gauge symmetries
\begin{align}\delta h_{\mu\nu}&=\partial_\mu\xi^T_\nu+\partial_\nu\xi^T_\mu+\chi\eta_{\mu\nu}, \nn\\
 \delta A_\mu&=-m\xi^T_\mu,\nn
\end{align}
where $\xi^{T}_{\mu}$ is transverse, $\partial^\mu\xi^{T}_{\mu}=0.$   The transversality of $\xi_\mu$ results from the constraint \eqref{Weylconstraint2} once $\xi$ has been eliminated.

It is instructive to see how the equations of a massive spin-2 field emerge from this Lagrangian. Taking the divergence of the $A_{\mu}$ equations of motion gives
\be
\partial^{\mu} \partial^{\nu} h_{\mu \nu} - \frac{1}{D} \Box h =0.
\ee
The $A_{\mu}$ equations and the divergence of the $h_{\mu \nu}$ equations then together imply
\be \label{xmas1}
\Box A_{\nu}  = -m\left(\partial^{\mu} h_{\mu \nu}-\frac{1}{D} \partial_{\nu} h \right), \quad \quad \partial_{\nu} \partial^{\alpha} A_{\alpha}  =0.
\ee
If we define
\be
\bar{h}_{\mu \nu} \equiv  h_{\mu \nu} + \frac{1}{m}\partial_{\mu} A_{\nu} + \frac{1}{m}\partial_{\nu} A_{\mu},
\ee
then the $h_{\mu \nu}$ equations give
\be \label{xmas2}
(\Box -m^2)\left(\bar{h}_{\mu \nu} - \frac{1}{D} \eta_{\mu \nu} \bar{h}\right)=0,
\ee
and \eqref{xmas1} gives
\be \label{xmas3}
\partial^{\mu} \bar{h}_{\mu \nu} - \frac{1}{D} \partial_{\nu} \bar{h} =0.
\ee
The field $\bar{h}_{\mu \nu}$ inherits the gauge transformation $\delta \bar{h}_{\mu \nu} = \chi \eta_{\mu \nu}$, so we can gauge fix $\bar{h} =0$. Equations \eqref{xmas2} and \eqref{xmas3} then show that $\bar{h}_{\mu \nu}$, a combination of the original tensor and vector, satisfy the equations of a massive spin-2 field.

This massive WTDiff theory can be obtained from the massive Fierz-Pauli theory \eqref{FPmassive} by the formal replacement
\be h_{\mu \nu} \rightarrow h_{\mu \nu} -\frac{1}{D}\eta_{\mu \nu} h + \frac{1}{m}\partial_{\mu} A_{\nu} + \frac{1}{m}\partial_{\nu} A_{\mu} .\ee
We emphasize that this is a not an invertible field redefinition or a St\"uckelberg transformation, just a formal replacement that we can make at the level of the action to generate the new theory.

\subsubsection{Gauge $A'_\mu=0$}
\textcolor{black}{If we instead use $\xi^\mu$ to gauge fix $A'_{\mu} \equiv A_{\mu} +\frac{D}{2m} \partial_{\mu} \phi=0$ in \eqref{mWTDiffminvector1}}, 
we are left with a scalar-tensor description of massive WTDiff. 
This gives
\begin{align} \label{mWTDiffMin}
{\cal L}_{{\rm WTDiff},D}(h, \phi) & = {\cal L}_{{\rm WTDiff},D}-{1\over 2}m^2\left(h_{\mu\nu}^2-{1\over D}h^2\right)  +(D-2)h^{\mu\nu}\left(\partial_\mu\partial_\nu\phi-{1\over D}\eta_{\mu\nu}\square\phi\right)\nn\\
&+{(D-1)\over 2}\left[(D-2)(\partial\phi)^2+D{m^2}\phi^2\right],
\end{align}
which has the residual gauge symmetries
\begin{align}  \delta h_{\mu\nu}&=\chi\eta_{\mu\nu}, \nn\\
\delta\phi&=0.\label{residweys}
\end{align}
Note that $\xi$ is not left as a residual gauge parameter because it was constrained in terms of $\xi_\mu$ by \eqref{Weylconstraint2}.

It is instructive to see how the equations of a massive spin-2 field emerge from this Lagrangian.  The $\phi$ equation of motion is
\be
(D-2) \Box \phi - D m^2 \phi =  \frac{(D-2)}{(D-1)} \left(\partial^{\mu} \partial^{\nu} h_{\mu \nu}-\frac{1}{D} \Box h \right),\label{scalareq1}
\ee
and $\partial^{\mu} \partial^{\nu}$ acting on the $h_{\mu \nu}$ equations of motion gives
\be
(D-2) \Box \left(\partial^{\mu} \partial^{\nu} h_{\mu \nu}-\frac{1}{D} \Box h \right)+m^2D \left(\partial^{\mu} \partial^{\nu} h_{\mu \nu}-\frac{1}{D} \Box h \right) =  (D-1)(D-2) \Box^2 \phi.\label{tenseq1}
\ee
Solving \eqref{scalareq1} for $\partial^{\mu} \partial^{\nu} h_{\mu \nu}-\frac{1}{D} \Box h$ and substituting it into \eqref{tenseq1}, we find that all the higher derivatives acting on $\phi$ cancel and we are left with
 $m^4 \phi=0$, which tells us that the scalar vanishes. 
 Using this in \eqref{scalareq1}, and in $\partial^{\mu}$ acting on the $h_{\mu \nu}$ equations of motion, then gives the constraint
\be
\partial^{\mu} h_{\mu \nu} - \frac{1}{D} \partial_{\nu} h =0,
\ee
and using it in the $h_{\mu \nu}$ equations gives
\be
(\Box -m^2) \left( h_{\mu \nu} - \frac{1}{D} \eta_{\mu \nu} h \right) =0.
\ee
Now we can use the symmetry \eqref{residweys} to choose the gauge $h=0$, and we recover the standard massive spin-2 equations \eqref{KG}.
Thus $\phi$ is not dynamical, but rather its equation of motion combines with those of the tensor to enforce the transversality constraint, $\partial^{\mu} (h_{\mu \nu} - \frac{1}{D} \eta_{\mu \nu} h) =0$. 
However, even though $\phi$ is ultimately nondynamical, we cannot directly integrate it out of the action because it appears with derivatives.

This theory \eqref{mWTDiffMin} is a Weyl St\"uckelberg-ed version of the Fierz-Pauli theory.  We can recover Fierz-Pauli by gauge fixing $h_{\mu\nu}$ to be traceless using the Weyl symmetry \eqref{residweys}, $h_{\mu\nu}\rightarrow \tilde h_{\mu\nu}$, where $\tilde h_{\mu\nu}$ is traceless,
\begin{align} \label{tracelessmassive}
{\cal L}_{{\rm traceless},D}=&-{1\over 2}\partial_\lambda \tilde{h}_{\mu\nu}\partial^\lambda \tilde{h}^{\mu\nu}+\partial_\lambda \tilde{h}_{\mu\nu}\partial^\mu \tilde{h}^{\lambda\nu} -{1\over 2}m^2\tilde{h}_{\mu\nu}^2+(D-2)\tilde{h}^{\mu\nu}\partial_\mu\partial_\nu\phi \nn \\ 
&+\frac{(D-1)}{2}\left[(D-2)(\partial\phi)^2+D{m^2}\phi^2\right],
\end{align}
and then repackaging $D \phi$ to be the trace of $h_{\mu\nu}$,
\be
h_{\mu \nu} \equiv \tilde{h}_{\mu \nu} + \eta_{\mu \nu} \phi,
\ee
recovering \eqref{FPmassive}. 
Going the other way, we can get \eqref{mWTDiffMin} from the Fierz-Pauli theory \eqref{FPmassive} by the replacement
\be h_{\mu \nu} \rightarrow h_{\mu \nu} -\frac{1}{D}\eta_{\mu \nu} h + \phi \eta_{\mu \nu} ,\ee
which is a Weyl St\"uckelberg transformation followed by a field redefinition of the scalar.

\subsection{Massive traceless}

Dimensionally reducing the traceless theory \eqref{tracelessm} gives
\begin{align} \label{massivetraceless}
\mathcal{L}_{\rm m\ traceless, D}=& -{1\over 2}\partial_\lambda h_{\mu\nu}\partial^\lambda h^{\mu\nu}+\partial_\lambda h_{\mu\nu}\partial^\mu h^{\lambda\nu}-{1\over 2}\partial_\mu h \partial^\mu h -\frac{1}{2}m^2(h_{\mu \nu}h^{\mu \nu} - h^2) \nn \\ & -\frac{1}{2} F_{\mu \nu}^2 -2m h^{\mu \nu}(\partial_{\mu} A_{\nu} - \eta_{\mu \nu} \partial^{\alpha} A_{\alpha}),
\end{align}
with the gauge transformations
\begin{align}\delta h_{\mu\nu}&=\partial_\mu\xi_\nu+\partial_\nu\xi_\mu \nn\\
 \delta A_\mu&=\partial_\mu\xi-m\xi_\mu,
\end{align}
and the constraint
\be
\partial^{\mu} \xi_{\mu} +m \xi =0.
\ee
Here, the lower-dimensional $h_{\mu\nu}$ is not traceless, but there is no scalar field present because of the requirement that the higher-dimensional trace vanish.
\textcolor{black}{This theory can be obtained from either the St\"uckelberg-ed massive Fierz-Pauli theory \eqref{mFPStueckelberg} or the  St\"uckelberg-ed massive WTDiff theory \eqref{mWTDiffminvector1} by using the $U(1)$ gauge symmetry $\xi$ to fix $\phi = h/D$.}
If we use the Weyl gauge freedom to fix $h=0$ in the massive WTDiff theory with the scalar gauged away, \eqref{mWTDiffvector}, then this is the same as \eqref{massivetraceless} after fixing $h=0$ using the $U(1)$, which we can do algebraically because of the constraint.

\subsection{Massive conformal scalar tensor}
As with the massless theories, we can find a parent theory which when gauge fixed gives either a massive Fierz-Pauli theory or massive WTDiff theory. Performing the KK reduction on the higher-dimensional conformal scalar-tensor theory \eqref{conformal} gives
\bea \label{conformalmassive}
{\cal L}_{{\rm Conformal},D+1} \rightarrow &&\mathcal{L}_{mFP, D} -\frac{1}{2} F_{\mu \nu}^2-2mh^{\mu \nu} \left( \partial_{\nu} A_{\mu}- \eta_{\mu \nu} \partial^{\alpha} A_{\alpha} \right) + h^{\mu \nu} \left( \partial_{\mu} \partial_{\nu} \phi - \eta_{\mu \nu} \Box \phi \right) \nn \\
&& + \frac{1}{2}D(D-1) \left( \partial_{\mu} \varphi \partial^{\mu} \varphi + m^2 \varphi^2 \right)+(D-1)h^{\mu\nu} \left( \partial_{\mu} \partial_{\nu} \varphi - \eta_{\mu \nu}(\Box - m^2)\varphi \right) \nn \\
&& -2(D-1)mA_{\mu} \partial^{\mu} \varphi + (D-1) \partial^{\mu} \phi \partial_{\mu} \varphi.
\eea
There are now two scalars: $\varphi$ coming from the dimensional reduction of the higher-dimensional parent scalar and $\phi$ from the $yy$ component of the graviton.
\textcolor{black}{Redefining $\phi \rightarrow -(D\phi + \varphi)$ then gives the massive conformal theory, as represented in Fig. \ref{plot2},
\begin{align} \label{conformalmassive2}
{\cal L}_{{\rm m\ Conformal},D} = & \mathcal{L}_{mFP, D} -\frac{1}{2} F_{\mu \nu}^2-2mh^{\mu \nu} \left( \partial_{\nu} A_{\mu}- \eta_{\mu \nu} \partial^{\alpha} A_{\alpha} \right) -D h^{\mu \nu} \left( \partial_{\mu} \partial_{\nu} \phi - \eta_{\mu \nu} \Box \phi \right) \nn \\
& + \frac{1}{2}(D-1)  (D-2)\partial_{\mu} \varphi \partial^{\mu} \varphi +(D-1)m^2\left( \frac{D}{2} \varphi^2 + h \varphi\right) \nn \\ &+(D-2)h^{\mu\nu} \left( \partial_{\mu}  \partial_{\nu} \varphi - \eta_{\mu \nu}\Box \varphi \right) -2(D-1)mA_{\mu} \partial^{\mu} \varphi -D (D-1) \partial^{\mu} \phi \partial_{\mu} \varphi,
\end{align}
which has the gauge symmetries
\begin{align} & \delta h_{\mu\nu} =\partial_\mu\xi_\nu+\partial_\nu\xi_\mu+\chi\eta_{\mu\nu}, \nn\\
&\delta A_\mu  =\partial_\mu\xi-m\xi_\mu,\nn\\
 &\delta\phi =-\frac{2}{D}m \xi, \nn \\
&\delta \varphi = - \chi.
\end{align}
}
To recover the St\"uckelberg-ed Fierz-Pauli theory \eqref{mFPStueckelberg} from \eqref{conformalmassive2}, we gauge fix
\be \varphi=0.\ee
Conversely, we can get the conformal theory from the St\"uckelberg-ed Fierz-Pauli theory by applying a Weyl St\"uckelberg transformation, 
\be h_{\mu \nu} \rightarrow h_{\mu \nu} + \varphi\, \eta_{\mu \nu}.\ee
To recover the St\"uckelberg-ed WTDiff theory \eqref{mWTDiffminvector1} from \eqref{conformalmassive2}, we gauge fix
\be \varphi=-{1\over D}h+\phi.\ee

\subsection{Massless limit of massive WTDiff}

Taking the massless limit of \eqref{mWTDiffminvector1} gives
\be \label{masslessmWTDiff}  {\cal L}_{{\rm WTDiff},D} -{1\over 2}F_{\mu\nu}^2 -2h^{\mu\nu}\left(\partial_\mu\partial_\nu\phi-{1\over D}\eta_{\mu\nu}\square\phi\right)-{(D-1)(D+2)\over 2}(\partial\phi)^2,
\ee
which has the gauge symmetries
\begin{align} \delta h_{\mu\nu}&=\partial_\mu\xi^T_\nu+\partial_\nu\xi^T_\mu+\chi\eta_{\mu\nu}, \nn\\
 \delta A_\mu&=\partial_\mu\xi,\nn\\
 \delta\phi&=0.
\end{align}
In the massless limit, we expect the $\frac{(D+1)(D-2)}{2}$ massive spin-2 degrees of freedom to decouple into a massless spin 0, a massless spin 1 and a massless spin 2.  The massless spin 1 decouples, but the scalar-tensor part contains cross terms that are not easily unmixed.\footnote{In the analogous situation for Fierz-Pauli, these cross terms are unmixed by a linear Weyl transformation, which is not possible here due to the Weyl symmetry of the kinetic term.}  Nevertheless, this scalar-tensor Lagrangian still describes massless helicity-2 and -0 modes.

The equations of motion for the scalar-tensor part of this theory can be written as
\be
\Box(h_{\mu \nu} - \frac{1}{D}\eta_{\mu \nu} h) = 2\left( \partial_{\mu} \partial_{\nu} \phi - \frac{1}{D} \eta_{\mu \nu} \Box \phi \right), \quad \quad
\ee
and
\be
\Box \phi = \frac{2}{(D-1)(D+2)}\left(\partial^{\alpha} \partial^{\beta} h_{\alpha \beta} - \frac{1}{D} \Box h \right),
\ee
with the constraint
\be
\partial^{\alpha} \partial^{\beta} h_{\alpha \beta} - \frac{1}{D} \Box h = c, 
\ee
for an arbitrary integration constant $c$. If we want a transverse-traceless tensor, then we need to pick the $c=0$ solution, in which case the scalar is unsourced. For nonzero $c$, the scalar-tensor coupling acts on shell like a constant source term for $\phi$, which is why we cannot unmix the fields in \eqref{masslessmWTDiff} without imposing an extra constraint.

The above massless limit of the St\"uckelberg-ed form of the theory preserves the number of degrees of freedom. However, we can also consider setting $m=0$ in \eqref{mWTDiffMin} or \eqref{mWTDiffvector}, which does not preserve the number of degrees of freedom. The analogous procedure for Fierz-Pauli gives back the correct massless spin-2 theory, linearized GR. For \eqref{mWTDiffvector}, setting $m=0$ gives a Lagrangian with a noninteracting vector and tensor. This has 4 propagating degrees of freedom, so we have lost 1 degree of freedom.  For \eqref{mWTDiffMin}, setting $m=0$ gives the following Lagrangian:
\be \label{WTDiff2}
 {\cal L}_{{\rm WTDiff},D}+(D-2)h^{\mu\nu}\left(\partial_\mu\partial_\nu\phi-{1\over D}\eta_{\mu\nu}\square\phi\right)+{(D-1)(D-2)\over 2}(\partial\phi)^2,
\ee
which has the gauge symmetries
\begin{align} \delta h_{\mu\nu}&=\partial_\mu\xi_\nu+\partial_\nu\xi_\mu+\chi\eta_{\mu\nu}, \nn\\
 \delta\phi&=\frac{2}{D}\partial^{\mu} \xi_{\mu}.
\end{align}
This theory is clearly different from \eqref{WTDiffmassless} and is in fact related to the conformal scalar-tensor theory \eqref{conformal} by $\phi=\varphi+h/D$.

The fact that we can set $\phi=0$ \textcolor{black}{by hand} in the Lagrangian \eqref{WTDiff2} and still have a Weyl invariant theory that propagates the same number of degrees of freedom is a peculiarity special to the massless theory. Indeed, if we by hand kill $\phi$ in \eqref{mWTDiffMin}, we introduce a ghost. 

\section{Curved-space theories} \label{sec:curvemassless}
In this section, for completeness, we generalize the results of Secs. \ref{sec:flatmassless} and \ref{sec:flatmassive} to curved space. Interesting new features, such as the existence of partially massless theories in the massive case, arise only in curved space.

In this section all indices are raised and lowered with the background metric ${g}_{\mu \nu}$ and all covariant derivatives and curvature tensors are those of ${g}_{\mu \nu}.$ All \textcolor{black}{linear} Lagrangians, e.g. $\mathcal{L}_{FP}$, refer to the curved-space versions, and we omit the overall factor of $\sqrt{-{g}}$.

As is well known, a massless graviton can only consistently propagate on an Einstein space, i.e. a space which solves the vacuum Einstein equations with cosmological constant $\Lambda$ \cite{Aragone:1979bm,Deser:2006sq}.  The background quantities therefore satisfy 
\be
R_{\mu \nu} = \frac{R}{D} {g}_{\mu \nu}, \quad \quad {\Lambda} = \frac{(D-2)}{2D}R.
\ee
For a massive graviton, there is no such restriction, and it can propagate on any background \cite{Bernard:2014bfa}.  However, for simplicity we will restrict to the Einstein case even for the massive actions.

To find curved-space massless actions, we minimally couple the corresponding flat-space action and then add curvature terms with coefficients chosen so as to maintain invariance under the gauge symmetries.
To find \textcolor{black}{curved-space} massive actions, we minimally couple the corresponding flat-space action and then add curvature terms in such a way that any relevant constraints necessary to preserve the correct number of degrees of freedom are preserved.  Alternatively, we can find curved-space massive actions in maximally symmetric backgrounds from higher-dimensional flat actions by performing a Kaluza-Klein reduction along a radial direction \cite{Biswas:2002nk}.

\subsection{Fierz-Pauli}

The massless Fierz-Pauli Lagrangian in curved space is
\be \label{curveFPmassless} \mathcal{L}_{\rm FP, D} = - \frac{1}{2} \nabla_{\alpha} h_{\mu \nu} \nabla^{\alpha} h^{\mu \nu} +\nabla_{\alpha} h_{\mu \nu} \nabla^{\nu} h^{\mu \alpha} - \nabla_{\mu} h \nabla_{\nu} h^{\mu \nu} + \frac{1}{2} \nabla_{\mu} h \nabla^{\mu} h + \frac{R}{D}(h_{\mu \nu}h^{\mu \nu} -\frac{1}{2} h^2).
\ee
This is the linearization of the Einstein-Hilbert Lagrangian with a cosmological constant, $\mathcal{L} = \frac{1}{2 \kappa^2} \sqrt{-g} (R-2{\Lambda})$, around an Einstein background \textcolor{black}{(so that the full metric is ${g}_{\mu \nu}+2 \kappa h_{\mu \nu}$)}.  It has the gauge symmetry 
\bea && \delta h_{\mu \nu} = \nabla_{\mu} \xi_{\nu} +\nabla_{\nu} \xi_{\mu}. \nn
\eea

To get massive spin 2 in curved space, we simply add the Fierz-Pauli mass term,
\be \label{mcurveFP} \mathcal{L}_{\rm mFP,D} = \mathcal{L}_{\rm FP,D}-\frac{1}{2} m^2(h_{\mu \nu}h^{\mu \nu} - h^2).
\ee
The divergence of the equations of motion gives the constraints
\be\label{divergenceconstraint}
\nabla^{\mu} h_{\mu \nu} - \nabla_{\nu}h = 0,
\ee
and the trace and double divergence combine to give
\be
h\left[ m^2(D-1) - \frac{R}{D}(D-2)\right] = 0.
\ee
If $R \not= m^2 \frac{D(D-1)}{(D-2)}$ then the trace is constrained,
$h=0,$ 
but if
\be \label{PMcondition}
R = m^2 \frac{D(D-1)}{(D-2)},
\ee
then the trace is unconstrained.
We consider these two cases in turn. 

\subsubsection{Generic massive spin 2}
When $R \neq m^2 \frac{D(D-1)}{(D-2)}$ we have the constraints
\be \label{transverse}
\nabla^{\mu} h_{\mu \nu} = 0,\ \ \ h = 0,
\ee
which when plugged into the equations of motion give the curved-space Klein-Gordon equation,
\be \label{curvemspin-2}
\left( \Box - m^2 \right)h_{\mu \nu}+ 2R_{\mu \rho\nu \sigma }h^{\rho \sigma}=0.
\ee
This describes $\frac{(D+1)(D-2)}{2}$ propagating degrees of freedom, the same as for a massive spin-2 field in flat space.

\subsubsection{Partially massless spin 2}

The special value \eqref{PMcondition} of the background curvature relative to the graviton mass is the special case in which a scalar gauge symmetry emerges,
\be
\delta h_{\mu \nu} = \nabla_{\mu} \nabla_{\nu} \alpha + \frac{m^2}{D-2}{g}_{\mu \nu} \alpha,
 \ee
for scalar gauge parameter $\alpha$,
 so that the resultant theory propagates 1 fewer degree of freedom than the generic massive graviton (for a total of 4 in four dimensions). This theory called partially massless \cite{Deser:1983mm} has been well studied at the linear level \cite{Higuchi:1986py,Brink:2000ag,Deser:2001pe,Deser:2001us,Deser:2001wx,Deser:2001xr,Zinoviev:2001dt,Deser:2004ji,Deser:2006zx,Deser:2013xb,Hinterbichler:2014xga} and attempts have been made to extend this to a nonlinear theory, although there are obstructions \cite{Zinoviev:2006im,Deser:2012qg,deRham:2013wv,Joung:2014aba,Zinoviev:2014zka,Garcia-Saenz:2014cwa}. The equations of motion in this case, after using the divergence constraint \eqref{divergenceconstraint}, are 
 \be
 \left( \Box - m^2 \right)h_{\mu \nu}+ 2R_{\mu \rho\nu \sigma }h^{\rho \sigma}- \left( \nabla_{\mu} \nabla_{\nu} + \frac{m^2}{D-2} {g}_{\mu \nu} \right) h = 0.
 \ee

\subsection{WTDiff}
The massless WTDiff Lagrangian \eqref{WTDiffmassless} extended to curved space is
\begin{align} \label{curvedWTDiff} {\cal L}_{{\rm WTDiff},D} =& - \frac{1}{2} \nabla_{\alpha} h_{\mu \nu} \nabla^{\alpha} h^{\mu \nu} +\nabla_{\alpha} h_{\mu \nu} \nabla^{\nu} h^{\mu \alpha} -\frac{2}{D} \nabla_{\mu} h \nabla_{\nu} h^{\mu \nu} + \frac{(D+2)}{2D^2} \nabla_{\mu} h \nabla^{\mu} h  \nn \\ &+ \frac{R}{D}(h_{\mu \nu}h^{\mu \nu} -\frac{1}{D} h^2).
\end{align}
This is determined by ensuring the gauge symmetry
\bea && \delta h_{\mu \nu} = \nabla_{\mu} \xi^T_{\nu} +\nabla_{\nu} \xi^T_{\mu} + \chi \eta_{\mu \nu}, \nn
\eea
where $\nabla^{\mu} \xi^T_{\mu} = 0$.

To find the curved-space \textcolor{black}{scalar-tensor} massive WTDiff Lagrangian, we follow Fig. \ref{massiverelationships} and make the replacement $h_{\mu \nu} \rightarrow h_{\mu \nu} - \frac{1}{D}{g}_{\mu \nu}h+ {g}_{\mu \nu}\phi $ in the massive curved-space Fierz-Pauli theory \eqref{mcurveFP}. This gives
\begin{align} \label{mWTDiffcurved} \mathcal{L}_{ {\rm mWTDiff}, D} = & {\cal L}_{{\rm WTDiff},D}  -\frac{1}{2}m^2 \left(h_{\mu \nu} h^{\mu \nu} - \frac{1}{D}h^2\right)  + (D-2)h^{\mu \nu} \left( \nabla_{\mu} \nabla_{\nu} \phi - \frac{1}{D} {g}_{\mu \nu} \Box \phi \right) \nn \\
& + \frac{(D-2)(D-1)}{2} \partial_{\mu} \phi \partial^{\mu} \phi +  \left( \frac{D(D-1)}{2} m^2 -\frac{(D-2)}{2} R\right) \phi^2.
\end{align}
As expected, this reduces to \eqref{mWTDiffMin} in the flat-space limit.

Combining the gradient of the equations of motion in the following way, we find a vector constraint, proportional to the mass, expected in a ghost-free theory,
\be \label{vecconst} \nabla^{\nu} \frac{\delta \mathcal{L}}{\delta h^{\mu \nu}}+\frac{1}{D}\nabla_\mu \frac{\delta \mathcal{L}}{\delta \phi} =m^2\left[(D-1)\nabla_{\nu} \phi -\nabla^{\mu} h_{\mu \nu} + \frac{1}{D} \nabla_{\nu} h  \right] \textcolor{black}{=0}.
\ee
Combining the $\phi$ equation of motion and $\nabla^{\mu} \nabla^{\nu}$ acting on the $h_{\mu \nu}$ equations of motion in the following combination, 
\be\label{noethercond}
\nabla^{\mu} \nabla^{\nu} \frac{\delta \mathcal{L}}{\delta h^{\mu \nu}}+ \left(\frac{1}{D} \Box +\frac{m^2}{D-2}\right) \frac{\delta \mathcal{L}}{\delta \phi} ,
\ee
we find the following scalar constraint:
\be \label{condition12}
m^2 \phi \left[ R - m^2 \frac{D(D-1)}{(D-2)} \right] = 0.
\ee
Here, as in the Fierz-Pauli case, we find two different possibilities depending on whether the graviton mass is at the partially massless point \eqref{PMcondition} or not.

For $R \neq m^2 \frac{D(D-1)}{(D-2)}$, the condition \eqref{condition12} implies that $\phi =0$. Using the Weyl invariance to gauge fix $h=0$, we then get the transverse condition \eqref{transverse} from the vector constraint \eqref{vecconst}, and we get the curved-space massive spin-2 equation \eqref{curvemspin-2} from the $h_{\mu\nu}$ equations of motion, so the theory describes an ordinary massive spin 2 on the curved background, as we expect. 

For $R = m^2 \frac{D(D-1)}{(D-2)}$ on the other hand, the condition \eqref{condition12} is identically satisfied,
so \eqref{noethercond} is identically zero which means the Lagrangian possesses a scalar Noether identity.
The associated gauge symmetry is the partially massless gauge symmetry in this language and reads
\be \label{PMgaugesymmetry}
\delta h_{\mu \nu} = \nabla_{\mu} \nabla_{\nu} \alpha, \quad \quad \delta \phi = \left( \frac{1}{D} \Box+\frac{m^2}{D-2} \right) \alpha,
\ee
for scalar gauge parameter $\alpha$.
Notice that \eqref{PMcondition} is precisely the value of the background curvature that kills the quadratic scalar term in \eqref{mWTDiffcurved}. We can see from \eqref{PMgaugesymmetry} that this must be so, since under a partially massless gauge transformation the scalar quadratic term contains the only algebraic term in $\alpha$.\footnote{\textcolor{black}{This is related to the observation that we can write the nonderivative terms in \eqref{mcurveFP} as a Weyl invariant mass term, $\frac{D}{2(D-2)}m^2\left(h_{\mu \nu}h^{\mu \nu}-\frac{1}{D}h^2 \right)$, when \eqref{PMcondition} is satisfied.}}
Using the vector constraint \eqref{vecconst}, the $h_{\mu\nu}$ equations of motion in the partially massless case can be written in the form
\be
\left( \Box - m^2 \right)\left(h_{\mu \nu}-\frac{1}{D}{g}_{\mu \nu} h \right)+ 2R_{\mu \rho\nu \sigma }\left(h^{\rho \sigma}-\frac{1}{D}{g}^{\rho \sigma} h \right) = D \left( \nabla_{\mu} \nabla_{\nu} \phi - \frac{1}{D} g_{\mu\nu} \Box \phi \right),
\ee
and the $\phi$ equation is identically satisfied after using \eqref{vecconst}.

\subsection{Conformal scalar tensor}

The field replacements described in Fig. \ref{massiverelationships} allow us to find the WTDiff theories from the corresponding Fierz-Pauli theories, bypassing the  parent conformal scalar-tensor theory. Nevertheless, we write down the curved-space versions of these parent theories both for completeness and because they may be of interest in their own right. 

We find the curved-space version of \eqref{conformal} by making the replacement $h_{\mu \nu} \rightarrow h_{\mu \nu} + \varphi \, {g}_{\mu \nu}$ in \eqref{curveFPmassless}. This gives
\begin{align} 
{\cal L}_{{\rm m\ Conformal},D} = & \mathcal{L}_{mFP, D} 
 + (D-2)h^{\mu\nu} \left( \nabla_{\mu}  \nabla_{\nu} \varphi - {g}_{\mu \nu}\Box \varphi \right)\nn \\ & +\frac{1}{2}(D-1)(D-2)\partial_{\mu} \varphi \partial^{\mu} \varphi -\frac{(D-2)}{D}R \left(\frac{D}{2}  \varphi^2 + h \varphi \right)  ,
\end{align}
which has linearized Weyl and unconstrained linearized diff gauge symmetries
\be
\delta h_{\mu\nu} =\nabla_\mu\xi_\nu+\nabla_\nu\xi_\mu+\chi{g}_{\mu\nu},\ \ \ \ \delta\varphi=-\chi.
\ee
This theory is the linearization of the conformally coupled Einstein scalar-tensor theory around a curved background.

According to Fig. \ref{massiverelationships}, we should be able to get the curved-space version of the massive conformally coupled scalar theory through the following replacement in $\mathcal{L}_{mFP}$:
\be \label{ccmreplacement}
h_{\mu \nu} \rightarrow h_{\mu \nu} + \frac{1}{m}\nabla_{\mu} A_{\nu} + \frac{1}{m}\nabla_{\nu} A_{\mu} + \frac{D}{m^2}\nabla_{\mu} \nabla_{\nu} \phi + \varphi g_{\mu \nu}.
\ee
This gives
\begin{align} \label{curvedconformalmassive}
{\cal L}_{{\rm m\ Conformal},D} = & \mathcal{L}_{mFP, D} -\frac{1}{2} F_{\mu \nu}^2-2mh^{\mu \nu} \left( \nabla_{\nu} A_{\mu}- {g}_{\mu \nu} \nabla^{\alpha} A_{\alpha} \right) -D h^{\mu \nu} \left( \nabla_{\mu} \nabla_{\nu} \phi - {g}_{\mu \nu} \Box \phi \right) \nn \\
& + \frac{1}{2}(D-1)(D-2)\partial_{\mu} \varphi \partial^{\mu} \varphi + \left( (D-1)m^2-\frac{(D-2)}{D}R \right)\left(\frac{D}{2}  \varphi^2 + h \varphi \right)\nn \\ 
&  +(D-2)h^{\mu\nu} \left( \nabla_{\mu}  \nabla_{\nu} \varphi - {g}_{\mu \nu}\Box \varphi \right)-2(D-1)mA_{\mu} \partial^{\mu} \varphi -D (D-1) \partial^{\mu} \phi \partial_{\mu} \varphi \nn \\
& +\frac{2R}{D}\left( A_{\mu}^2 + \frac{D^2}{4m^{2}} \partial_{\mu} \phi \partial^{\mu} \phi +\frac{D}{m} A^{\mu} \partial_{\mu} \phi \right),
\end{align}
which has the gauge symmetries
\begin{align}  \delta h_{\mu\nu} &=\nabla_\mu\xi_\nu+\nabla_\nu\xi_\mu+\chi{g}_{\mu\nu}, \nn\\
\delta A_\mu & =\partial_\mu\xi-m\xi_\mu,\nn\\
 \delta\phi &=-\frac{2}{D}m \xi, \nn \\
\delta \varphi &= - \chi.
\end{align}
This is the curved-space version of \eqref{conformalmassive2}.

\section{Coupling to matter}

The various types of kinetic terms diff and WTDiff yield the same \textcolor{black}{(gauge-fixed)} equations of motion, \textcolor{black}{up to an integration constant}, since they are just different ways of describing the same propagating degrees of freedom. When coupling to matter, however, there can be differences in the allowable energy-momentum tensor to which the spin-2 field couples.\footnote{See \cite{Alvarez:2006uu} for a discussion of matter coupling in massless Fierz-Pauli and WTDiff in terms of propagators.}

\subsection{Massless theories}

The standard way to couple the massless Fierz-Pauli theory to a fixed external source $T^{\mu\nu}$ is by adding a term $\kappa \, h_{\mu \nu} T^{\mu \nu}$ to \eqref{FPmassless}, with $\kappa$ the gravitational coupling.  For the resultant theory to respect the gauge symmetry \eqref{lindifff}, we require that the source $T_{\mu \nu}$ is conserved, $\partial^{\mu} T_{\mu \nu}=0$. Choosing harmonic gauge, $\partial^{\mu} h_{\mu \nu} = \frac{1}{2} \partial_{\nu} h$, then gives 
\be
\Box h_{\mu \nu} = -\kappa\left(T_{\mu \nu} - \frac{1}{(D-2)} \eta_{\mu \nu} T\right),
\ee
\textcolor{black}{where $T \equiv T_{\mu}^{\mu}$. This is} the form of the sourced massless spin-2 equations as usually presented. For $T \neq 0$, we cannot find solutions satisfying both $h=0$ and the harmonic gauge choice, so this is not a good gauge choice for comparison with WTDiff. If we instead gauge fix $h=0$ but not harmonic gauge, then the equations of motion are
\be \label{FPmatter}
\Box h_{\mu \nu} - \partial_{\mu} \partial^{\alpha} h_{\alpha \nu}- \partial_{\nu} \partial^{\alpha} h_{\alpha \mu} = -\kappa \left(T_{\mu \nu} - \frac{1}{(D-2)} \eta_{\mu \nu} T\right),
\ee 
with the residual gauge invariance $\delta h_{\mu \nu} = \partial_{\mu} \xi^{T}_{\nu}+\partial_{\nu} \xi^{T}_{\mu}$. This is now in a form directly comparable to WTDiff.

We couple the massless WTDiff theory to a fixed external source $\tilde{T}^{\mu \nu}$ by adding a term $\kappa h_{\mu \nu} \tilde{T}^{\mu \nu}$ to \eqref{WTDiffmassless}. For the resultant theory to satisfy the gauge symmetry \eqref{gaugesymm1}, we require that
\be
\partial^{\mu} \tilde{T}_{\mu \nu} = \partial_{\nu} \Phi, \quad \tilde{T} =0,
\ee
for some scalar field $\Phi$, where $\tilde{T} \equiv \tilde{T}_{\mu}^{\mu}.$ These conditions are satisfied by
\be \label{tracelessT}
\tilde{T}_{\mu \nu} = T_{\mu \nu} - \frac{1}{D} \eta_{\mu \nu} T,
\ee
where $T_{\mu \nu}$ is \textcolor{black}{a conserved source}, $\partial^{\mu} T_{\mu \nu} =0$. Coupling $h_{\mu \nu}$ to \eqref{tracelessT} is also suggested by the horizontal line in Fig. \ref{plot1}, and this is what we consider henceforth.\footnote{We could also add to $\tilde{T}_{\mu \nu}$ terms of the form $(\partial_{\mu} \partial_{\nu} - \frac{1}{D} \eta_{\mu \nu} \Box) \theta$ for any scalar $\theta$, corresponding to the terms of the form $(\partial_{\mu} \partial_{\nu} - \eta_{\mu \nu} \Box) \theta$ that can be added to \textcolor{black}{the source} in the Fierz-Pauli theory.  These are the identically conserved ``improvement" terms.} This coupling implies that contributions to the \textcolor{black}{conserved source} that are a constant multiple of the background metric, i.e. cosmological constant contributions, do not appear in the massless Weyl invariant action. However, as we will now see, contributions proportional to the background metric reappear through an arbitrary integration constant. First, fix $h=0$ using the Weyl symmetry. Then, acting on the $h_{\mu \nu}$ equations of motion with $\partial^{\mu}$ gives, after integrating, 
\be
\partial^{\alpha} \partial^{\beta} h_{\alpha \beta} = -\frac{\kappa}{(D-2)}(T + c),
\ee
where $c$ is an arbitrary integration constant. Plugging this back in the $h_{\mu \nu}$ equations of motion gives
\be
\Box h_{\mu \nu} - \partial_{\mu} \partial^{\alpha} h_{\alpha \nu}- \partial_{\nu} \partial^{\alpha} h_{\alpha \mu} = -\kappa \left(T_{\mu \nu} - \frac{1}{(D-2)} \eta_{\mu \nu} \left(T+\frac{2c}{D}\right)\right),
\ee 
which has the residual gauge invariance $\delta h_{\mu \nu} = \partial_{\mu} \xi^{T}_{\nu}+\partial_{\nu} \xi^{T}_{\mu}$. Comparing to \eqref{FPmatter}, we see that this is the same as Fierz-Pauli except that in WTDiff energy-momentum contributions proportional to the background appear through an arbitrary integration constant. This is the linearized version of what happens to the cosmological constant \textcolor{black}{and vacuum energy contributions to the energy-momentum tensor} in unimodular gravity.

\subsection{Massive theories}

If we couple massive Fierz-Pauli to matter by adding a term $\kappa\, h_{\mu \nu} T^{\mu \nu}$ to \eqref{FPmassive}, then there is no gauge invariance to mandate that $T_{\mu \nu}$ be conserved. Nevertheless, it is simpler (and physically reasonable) to assume a conserved source, so we proceed with this assumption. The equation of motion is then\footnote{This implies the full equations of motion if we choose boundary conditions (e.g. retarded boundary conditions) such that $(\Box -m^2)f=0 \implies f=0$, as discussed in, e.g., \cite{Hinterbichler:2011tt}. We also assume this below.}
\be \label{mFPmatter}
(\Box -m^2) h_{\mu \nu} = - \kappa \left( T_{\mu \nu} - \frac{1}{(D-1)} \left(\eta_{\mu \nu} - \frac{1}{m^2} \partial_{\mu} \partial_{\nu} \right) T\right).
\ee

\textcolor{black}{In the massive WTDiff theory, we can know about the trace of the energy-momentum tensor by coupling to the auxiliary field, unlike in the massless case.} Indeed, such a coupling naturally arises when we apply the field replacement $h_{\mu \nu} \rightarrow h_{\mu \nu} - \frac{1}{D} \eta_{\mu \nu} + \phi \eta_{\mu \nu}$ to the term $h_{\mu \nu} T^{\mu \nu}$, so we expect the massive WTDiff theory to describe the same physics when coupled to matter as the massive Fierz-Pauli  theory. If we add to the Lagrangian \eqref{mWTDiffMin} the terms
\be
\Delta \mathcal{L}  = \kappa h_{\mu \nu} \left( T^{\mu \nu} - \frac{1}{D} \eta^{\mu \nu} T \right) + \kappa \phi T,
\ee
and assume a conserved source, \textcolor{black}{then the equations of motion for $h_{\mu \nu}$ and $\phi$ can be combined to give}
\be \label{mWTDiffmatter1}
\phi = -\frac{\kappa}{m^2D(D-1)}T .
\ee
The scalar is thus completely determined in terms of the source and hence is again nondynamical. The equations for $h_{\mu \nu}$ can then by given by 
\be \label{mWTDiffmatter2}
(\Box -m^2) \left(h_{\mu \nu}-\frac{1}{D} \eta_{\mu \nu} h \right) = - \kappa \left( T_{\mu \nu} - \frac{1}{D} \eta_{\mu \nu} T+\frac{1}{m^2(D-1)} \left( \partial_{\mu} \partial_{\nu} -\frac{1}{D}\eta_{\mu \nu} \Box \right) T\right).
\ee
Equations \eqref{mWTDiffmatter1} and \eqref{mWTDiffmatter2} together imply the full equations of motion, assuming appropriate boundary conditions and a conserved source; \textcolor{black}{if we identify $\phi$ with $h/D$}, then these are equivalent to the trace and traceless components of \eqref{mFPmatter}, respectively, so that the massive WTDiff theory is \textcolor{black}{physically equivalent to} massive Fierz-Pauli even when coupled to matter. The reason for this is clear; in the massless case, there was no scalar field and so contributions to the source proportional to the background could not couple, but the massive theory requires an auxiliary field to propagate the correct number of spin-2 degrees of freedom and this auxiliary field can couple to cosmological constant contributions. Of course, what matters is the coupling of these spin-2 fields to other degrees of freedom; \textcolor{black}{these degrees of freedom will not know about a cosmological constant if they are coupled only to $h_{\mu \nu}$, but they will if we allow couplings to the scalar field.}
 
\section{Discussion} \label{sec:discussion}

We have found linear Lagrangians that describe a massive spin-2 particle using a kinetic term that is locally Weyl and transverse-diffeomorphism invariant. This extends to the massive case the second of two theories of a massless spin-2 Lagrangian built from a symmetric rank-2 tensor field.  We found that some form of auxiliary field is a necessary ingredient in writing down such theories, and that these naturally arise from a Kaluza-Klein construction. 

Moreover, using the fact that the Fierz-Pauli and WTDiff theories can be viewed as different gauge fixings of a conformally coupled scalar-tensor theory, we were readily able to find the curved-space versions of these theories, bypassing a higher-dimensional construction. We discussed in detail the partially massless theories in this description.

Interestingly, the massless limit of the massive WTDiff theory defined with an auxiliary scalar differs from the usual massless WTDiff theory, giving instead a theory related to the conformal scalar-tensor theory by a field redefinition.

Many of these results extend nonlinearly (see \cite{Dengiz:2011ig, Tanhayi:2011aa} for another nonlinear Weyl invariant theory containing massive spin 2). The nonlinear version of WTDiff is a Weyl invariant version of unimodular gravity discussed in \cite{Alvarez:2006uu}. There should be a massive version of this nonlinear WTDiff theory, analogous to the de Rham-Gabadadze-Tolley (dRGT) ghost-free massive gravity that is built on the Einstein-Hilbert kinetic term. This should be fully equivalent to dRGT in terms of locally propagating degrees of freedom, but may differ in global degrees of freedom and matter couplings. 
Another possible direction to explore is the massive generalization of the TDiff scalar-tensor theories discussed in \cite{Alvarez:2006uu} and elsewhere.

{\bf Acknowledgements:} The authors would like to thank Tessa Baker, Clare Burrage, Macarena Lagos, John March-Russell, Johannes Noller, James Scargill, and Hans Winther for helpful discussions. J.B. is supported by the Rhodes Trust. P.G.F. acknowledges support from  STFC, BIPAC, a Higgs visiting fellowship and the Oxford Martin School. Research at Perimeter Institute is supported by the Government of Canada through Industry Canada and by the Province of Ontario through the Ministry of Economic Development and Innovation. This work was made possible in part through the support of a grant from the John Templeton Foundation. The opinions expressed in this publication are those of the authors and do not necessarily reflect the views of the John Templeton Foundation (K.H.). J.B. would like to thank the Perimeter Institute for hospitality while much of this work was completed.

\bibliographystyle{utphys}
\addcontentsline{toc}{section}{References}
\bibliography{arxivV3.bib}

\end{document}